\documentstyle[aps]{revtex}
\begin{document}
\draft
\title{Space Rotation Invariance}
\author{Andrey V. Novikov-Borodin \footnote{{\it e-mail addresses:}~~ novikov@al20.inr.troitsk.ru,~~ andrey.novikov-borodin@desy.de}}
\address{ Institute for Nuclear Research, Russian Academy of Sciences \\
117312 Moscow, 60-th October Anniversary prospect 7a, Russia}
\date{\today}
\maketitle
\begin{abstract}
The space rotation invariance hypothesis is examined. The basic space-time properties and the physical object description from this point of view are considered. An $\omega$--invariance as an approximation of the space rotation invariance hypothesis is introduced. It is shown that on frames of the $\omega$--invariance it is possible to describe the ``wave'' properties of elementary particles and to get the basic quantum mechanics equations, such as Schr{\"o}dinger and Klein-Gordon-Fock equations. The correlation between the space rotation objects and models of the elementary particles, quarks and even nuclei is found. The problems of metrics and gravitation from the space rotation invariance point of view are discussed. The introduced hypothesis may be a foundation of the theory of the Unification.  
\end{abstract}

%\thanks{* }

% body of paper here
\section*{Introduction}

A homogeneous, isotropic space and a homogeneous time were considered in Newton's physics. A homogeneous means the physics law invariance related to the translations in space and time. Isotropic means the physics law invariance related to the rotation about some fixed axis in space through some given angle. In other words, the Newton's physics was based on the supposition of the physics law invariance in the translated or rotated to each other frames of references in space at any time. \\ 

The special theory of relativity is based on the supposition of the physics law invariance of the {\it moving} straightly in space with constant velocity (inertial) frames of references. Some invariant ``formulae'' exist in these different frames, so we can use some references to correlate different frames to each other. The same invariance principle is used for local space regions in Einstein's general relativity theory. \\ 

On this way it seems to be logic to make the following step and suppose the physics law invariance in the {\it rotating} frames of references. Only the space rotation with the constant rotation frequency will be considered in this paper. The question is what can we use as a reference between frames? May be, the physical object in one frame of references can be interpreted as another physical object in other frame, and some known physical laws will be valid in both frames, but it will be {\it different} physical laws for {\it different} physical objects. In other words, may be, the meaning of ``invariance'' will be changed.  \\ 

So, we will try to investigate this question in this paper. We will try to define the space rotation in time and to get some, important for us, properties. Also we need to find a method of description of the physical objects in the rotation frames, to analyze types of physical objects and to find correlation with known physical objects and laws with space rotation objects and methods of their description. 

\section{Types of Space Rotation}
\label{sec:Types}

Let's consider a ``mathematical'', homogeneous, isotropic, three-dimensional space. \\

{\bf Axis Space Rotation.} 
We will call the space rotation (SR) of the $K'$ frame of references related to $K$ in time with the constant frequency $\omega$ about some fixed space axis as an {\it axis} space rotation (ASR). The new coordinates in $K'$ will be marked as $(x',y',z',t)=(X',t)$. The coordinates transformation between the ``initial'' frame $K$ and rotating one $K'$ will be written as: 

\begin{equation}
X'=X\cdot A^{\pm}_{z} =(A^{\pm}_{z})^{T} \cdot X^{T},\qquad A^{\pm}_{z}=
\left (\begin {array}{ccc} \cos(\omega\,t)&\mp\sin(\omega\,t)&0
\\\noalign{\medskip}\pm\sin(\omega\,t)&\cos(\omega\,t)&0
\\\noalign{\medskip}0&0&1\end {array}\right ).
\label{ASR}
\end{equation}

Here $X^{T}$ is a vector transposed to $X$ and $A^{\pm}_{z}$ is a transformation matrix of the ASR about $z$-axis in space (without loosing the generality). At any time the functional determinant of the ASR transformation matrix is not equal to zero (${\it det}A \neq 0$). We will call the transformation matrix normalized if its functional determinant is equal to one (${\it det}A=1$).  \\ 

For the empty space it is difficult to find any difference between initial $K$ and rotating $K'$ frames of references, because from the $K'$ point of view the initial one ($K$) will be the rotating frame with the same frequency, but in the opposite direction. So, for the ASR the following will be true: 

\begin{equation}
\left ( A^{\pm}\right )^{-1} = A^{\mp} = \left (A^{\pm}\right )^{T}. 
\label{n2}
\end{equation}

Here, $A^{-1}$ is the matrix, reversed to $A$.\\ 

For some fixed space point we will define a combination of rotations as a supposition of two or more ASR. Two different types of supposition take place, that corresponds to summation or multiplication of the transformation matrices. We will understand a product and a sum of ASR matrices $A_{1},A_{2}$ as usual: 
\[ A_1 \cdot A_2=\sum _{k=1}^{n}a^{(1)}_{ik}a^{(2)}_{kj}=A=\{ a_{ij}\},\qquad A_{1}+A_{2}=\{ a^{(1)}_{ij}+a^{(2)}_{ij}\}=A=\{a_{ij}\}. \]

Two or more ASRs will be called orthogonal or parallel to each other if their axes of rotation are orthogonal or parallel correspondingly. The product of two parallel ASRs ($\xi$ is a rotation axis, $\omega_{i}$ are the frequencies of rotation) is a ASR with the frequency $\omega=\omega_{1}+\omega_{2}$: 

\begin{equation}
A^{ASR}_{\xi}(\omega_{1})\cdot A^{ASR}_{\xi}(\omega_{2})= A^{ASR}_{\xi}(\omega_{1}+\omega_{2}).   
\label{omsum}
\end{equation}

The ASR transformation matrix can be normalized, because its functional determinant does not depend on time. \\

In equation (\ref{ASR}), the transformation matrix expression for the ASR about $z$-axis is presented, but it is simple to get the transformation matrix for the ASR about any arbitrary space axis, by usual coordinate transformation. In general, one need to make two axis space rotations $P^{ASR}_{1}$ and $P^{ASR}_{2}$ about coordinate axes by fixed angles $\phi_{1}$ and $\phi_{2}$ correspondingly for this coordinate transformation. The axis space rotation by fixed angle is a particular case of the considered ASR, so the transformation matrix for any ASR can be represented as a product of two fixed angle ASR transformation matrices and one ASR transformation matrix and all ASRs are about the coordinate axes: 

\begin{equation}
A^{ASR}=\, P^{ASR}_{x}\, P^{ASR}_{y}\, A^{ASR}_{z}.
\label{exp}
\end{equation}
  
We will call this expression as an {\it ASR coordinate expansion}. \\

Considering a set of all ASRs about all space axes at some fixed space point, one will conclude that the space rotation invariance (if it exists) has a local character. The initial or basic frame of references has meaning if and only if a {\it system} of physical objects is considered. \\ 

{\bf Multiple Space Rotation.} We will call the combination of ASR rotations with transformation matrices $A_1,A_2,\dots,A_n$ by definition as a {\it multiple} space rotation (MSR), if the general transformation matrix of this rotation $A^{MSR}$ will be expressed as: 

\begin{equation}
A^{MSR}=\, \prod _{k=1}^{n}A_{{k}}.
\label{MSR}
\end{equation}

Let the $K$ frame rotates related to $K'$ with the transformation matrix $A_{1}$ and $K''$ frame rotates related to $K$ with the transformation matrix $A_{2}$. The transformation between $K''$ and $K$ will be expressed by the transformation matrix $A^{MSR}=\{a_{ij}\}=A_1 \cdot A_2=\sum _{k=1}^{n}a^{1}_{ik}a^{2}_{kj}$.\\ 

The following properties for the MSR are true:\\ 
1. The reversed matrix is equal to transpose one:
\[\left ( A^{MSR}\right )^{-1} = \left (A^{MSR}\right )^{T}.\]
2. If all transformation matrices of the ASRs are normalized, the MSR transformation matrix is normalized too: 
\[ {\it det}A^{MSR} =\, \prod _{k=1}^{n} {\it det} A_{{k}} =1.\]
3. The permutation of any two ASRs in the MSR changes the resulted MSR transformation matrix:
\[A^{MSR}_{..ij..}=A_{1}\cdot A_{2}\cdot\dots\cdot A_{i}\cdot A_{j}\cdot\dots\cdot A_{n}\neq
A_{1}\cdot A_{2}\cdot\dots\cdot A_{j}\cdot A_{i}\cdot\dots\cdot A_{n}=A^{MSR}_{..ji..}.\] 
\\

We can consider the resulted matrix (\ref{exp}) for the ASR coordinate expansion as a particular case of the MSR, so all conclusions and MSR properties are valid for this expansion. Also, we can consider that any MSR can be represented by the combination of the ASRs about the coordinate axes. \\

{\bf Sum Space Rotation.} We will call by definition the combination of SR rotations with transformation matrices $A_1,A_2,\dots,A_n$ as a {\it sum} space rotation (SSR) if their general transformation matrix $A^{SSR}$ of the SR of $K'$ related to $K$ will consist of the sum transformation matrices: 

\begin{equation}
A^{SSR}=\, \sum _{k=1}^{n}\, A_{{k}}.
\label{SSR}
\end{equation}

These SRs are considered from the $K$ frame of references. \\

For the SSR the following properties are true: \\
1. The reversed matrix is not equal to transpose one:
 \[\left ( A^{SSR}\right )^{-1} \neq \left (A^{SSR}\right )^{T}.\]
2. In general, even if any transformation matrix of ASR is normalized, the SSR transformation matrix is not normalized: 
\[ {\it det}A^{SSR}\neq 1.\]
3. The permutation of any two SR in SSR does not change the resulted SSR transformation matrix:
\[A^{SSR}_{..ij..}=A_1+A_2+\dots+A_i+A_j+\dots+A_n=
A_1+A_2+\dots+A_j+A_i+\dots+A_n=A^{SSR}_{..ji..}.\] 

Due to considered matrix properties, we can conclude that the transformation matrix of {\it any space rotation} may be represented as a sum of products of the ASR transformation matrices: 

\begin{equation}
A^{SR}=\, \sum_{i}\, A^{MSR}_{i}. 
\label{SR}
\end{equation}

\section{Quantum Mechanics Equations}
\label{sec:Quantum}

Let $K'$ is a ASR frame related to $K$ with transformation matrix $A^{ASR}$. Let's fix some two events in $K$: $(X_{1},t_{1})$ and $(X_{2},t_{2})$. As far as the matrix $A^{ASR}$ is only time dependent $A^{ASR}=A^{ASR}(t)$, for $X'_{2}-X'_{1}$ in $K'$ one can get from definition (\ref{ASR}): 

\begin{equation}
X'_{2}-X'_{1}=\, X_{2}A^{ASR}(t_{2})-X_{1}A^{ASR}(t_{1}).
\label{n5}
\end{equation} 
  
The matrix $A^{ASR}$ is periodical with the period $T=2\pi/\omega$. So for time points  $t_{2}=t_{1}+2\pi k/\omega$, where $k$ is integer, $A^{ASR}(t_{1}+2\pi k/\omega)=A^{ASR}(t_{1}) =A^{ASR}_{|2\pi k/\omega}$ and from (\ref{n5}) we will get: 

\begin{equation}
\Delta X'_{|2\pi k/\omega}=\,  \Delta X \cdot A^{ASR}_{|2\pi k/\omega}.
\label{n6}
\end{equation}

If the ASR matrix is normalized (${\it det}A=1$): 

\begin{equation}
\|\Delta X'\|^{2}_{|2\pi k/\omega}=\, \|\Delta X\|^{2}. 
\label{n7}
\end{equation}

This means that in some time points $t_{2}=t_{1}+2\pi k/\omega$ the equality (\ref{n7}) is true and consequently the interval $ds^{2}=c^{2}dt^{2}-dx^{2}-dy^{2}-dz^{2}$ as it was defined in Minkowski space is invariant in $K$ and $K'$. Time points $t_{2}=t_{1}+2\pi k/\omega$ create the numerable infinite aggregate on the $t$-axis. The rotation frequency $\omega$ is the initial parameter of this aggregation. We will call by definition that rotating frames $K$ and $K'$ are {\it $\omega$ - invariant}. This means that frames on this infinite aggregate are Lorenz-invariant. As it was mentioned before, this invariance is based on the numerable infinite aggregate on $t$-axis, so the frequency $\omega$ defines the {\it scale} between two frames of references $K$ and $K'$. On this approach, this parameter seems and needs to be very important in the physical object description in these frames. \\ 

On this approach, we can get the basic quantum mechanics equations.
If a physical object (some its property) is described in $K'$ by the function $q(X',\tau)$ ($\tau = ct$, c is a speed of light), in $K$, due to the $\omega$ - invariance, in time-invariant points one can define the corresponding function $\psi(X,\tau)$ ({\bf i} is an imaginary unit) as:

\begin{equation}
\psi(X,\tau)_{|2\pi k/\omega}=q(X',\tau)\exp(\pm {\bf i}\Omega\tau),  
\label{SRpsi}
\end{equation} 

because $\exp(\pm {\bf i}\Omega\tau)_{|2\pi k/\omega}=1$ and in these points $X'_{|2\pi k/\omega}=X$. Here, we assign $\Omega=\omega/c$. Also, if we consider, that the physical object in $K'$ is stable, or, at least, stable in comparison with the period of rotation $T=2\pi /\omega$, the previous expression  can be written as: 

\begin{equation}
\psi(X,\tau)_{|2\pi k/\omega}=q(X)\exp(\pm {\bf i}\Omega\tau).
\label{ps}
\end{equation} 

Furthermore, we will replace the numerable infinite aggregate on $t$-axis by the real $t$-axis. At any, even very high values of $\omega$, this approximation may be quite accurate, but always not complete. \\

Following \cite{NB99}, we will consider this physical object (\ref{ps}) in the inertial frame $K^{in}$ related to $K$. The invariant coordinate transformations between $K$ and $K^{in}$ in special theory of relativity are given by Lorentz transformations: 

\begin{equation}
x=x^{in},\qquad y=y^{in},\qquad z=\gamma(z^{in}-\beta\tau^{in}),\qquad \tau=\gamma(\tau^{in}-\beta z^{in}),
\label{Lor}
\end{equation}

where $\beta=V/c$, $\gamma=1/\sqrt{1-\beta^2}$, $V$ is a speed of the $K^{in}$ frame related to $K$, and, without loosing the generality, we consider that the $K^{in}$ frame is moving parallel to $z$- and $z^{in}$-axes. \\

In the frame $K^{in}$, taking into account (\ref{Lor}) and designating $\xi=\gamma(z^{in}-\beta\tau^{in})$, $\eta=\gamma(\tau^{in}-\beta z^{in})-\tau^{in}$, the considered function (\ref{ps}) may be represented as:

\begin{equation}
\psi^{in}(x,y,z^{in},\tau^{in})=q(x,y,\xi)\exp[{\bf i}\Omega^{in}(\eta+\tau^{in})].
\label{psiin}
\end{equation}

Omitting the $^{in}$ indexes and $\pm$ for the notation simplicity, we will represent (\ref{psiin}) as follows: 

\begin{equation}
\psi(X,\tau)=\psi^b(X,\tau)\exp({\bf i}\Omega\tau),\qquad 
\psi^b(X,\tau)=q(x,y,\xi)\exp({\bf i}\Omega\eta).
\label{pso}
\end{equation}

It will be shown later that on these assignments the function $\psi^b(X,\tau)$ gets the meaning of the de Broglie's wave. Combining the first and second partial derivatives by space and time coordinates for the function $\psi^b(X,\tau)$, it is possible to cancel imaginary parts and to get the equation: 

\begin{equation}
-i\gamma{1\over\psi^b}{\partial\psi^b\over\partial\tau}+
{1\over 2\Omega\psi^b}\nabla^2\psi^b={1\over 2\Omega}{\nabla^2q\over q}+
{\Omega\over 2}(\gamma-1)^2.
\label{schrod}
\end{equation} 

Here $\nabla^2\equiv\partial^2/\partial x^2+\partial^2/\partial y^2+
\partial^2/\partial z^2$ is a Laplace operator. The second term in the right part of the equation (\ref{schrod}) speeds to zero fast enough $(\sim\beta^4)$ with non-relativistic velocities. If it is possible to separate variables in some frame of references or, at least, to separate the time variable in the function $q$, that corresponds to some stable or quasi-stable states, so $\nabla^2 q/q=\, u(x,y,z)$ and, supposing $\Omega=mc/\hbar$, where $m$ and $\hbar$ are understood as a rest mass and a Planck constant, we will get: 

\begin{equation}
-i\hbar c\gamma{\partial\psi^b\over\partial\tau}+
{\hbar^2\over 2m}\nabla^2\psi^b=\left[{\hbar^2\over 2m}u(x,y,z)+
mc^2{(\gamma-1)^2\over 2}\right]\psi^b.
\label{Schrodinger}
\end{equation}

Designating $\hbar^2u(x,y,z)/2m=U(x,y,z)+E$, where $E$ is some constant,
we will get in passage to the limit $(\gamma\rightarrow 1)$ (in non-relativistic
case) the {\it Schr\"{o}dinger} equation. Usually, the function $U(x,y,z)$ is understood as an external potential function, $E$ has meaning of the energy of the object. As it was mentioned before, $\psi^b(x,y,z,\tau)$ agrees with the {\it de Broglie's} description of the particle wave properties. \\
If the potential function $U(x,y,z)\rightarrow 0$ with $\| X\| \rightarrow\infty$, so the constant $E$ has a meaning of a kinetic energy of the ``free''
physical object in a fundamental frame of references of the force field. So, for
the attractive (for the object) field, $E<0$ corresponds to the case of the
capture of the object by the field and some stable states of the system may
exist, $E\geq 0$ corresponds to a free motion of the object. \\
Also, in distinguish with the Schr\"{o}dinger equation, the equation (\ref{schrod})
and, with some stipulations, (\ref{Schrodinger}) have to be true also in the relativistic
case. \\

Combining the second partial derivatives by space and time coordinates for the function $\psi(X,\tau)$ (\ref{pso}), it is also possible to cancel imaginary parts and to get another equation:

\begin{equation}
{1\over\psi}\left({\partial^2\psi\over\partial\tau^2}-\nabla^2\psi\right)=
-\left({\nabla^2 q-\beta^2 \partial^2 q/\partial z^2\over q}+\Omega^2\right).
\label{Klein}
\end{equation}

There are included functions only from space coordinates in $K$ on the right 
part of the equation ($[\nabla^2 q^{in}-\beta^2 \partial^2 q^{in}/\partial (z^{in}) ^2]/q^{in}=\nabla^2 q/q$),
but on the left part - in frame $K$. Thus, with the stability condition of
the considered physical object, it needs to be a scalar on the right part of (\ref{Klein})
in brackets. Designating this scalar as $(mc/\hbar)^2$, we will get the
{\it Klein-Gordon-Fock equation} for a free relativistic (pseudo-) scalar
particle with rest mass $m$, that corresponds to the standard model,
when the plane monochromatic wave is confronted to the particle (without
spin). Thus, the introduced functions (\ref{pso}) may be interpreted as a wave
function of the physical object. \\ 

In the particular case, the equation (\ref{Klein}) with the zero scalar on the right part corresponds to the {\it wave equation}.\\

It is considered in quantum physics that the $t$-axis is continuous. From our point of view, in quantum physics the numerable infinite aggregate on $t$-axis is replaced by the continuous $t$-axis. At any, even very high values of $\omega$, this approximation may be quite accurate, but always not complete. On this approach, it becomes clear that the $\omega$ - invariance is the reason of the {\it uncertainties} in quantum mechanics, the reason of its incompleteness and formalism \cite{Mes}. M. Gell-Mann \cite{G-M} characterized the quantum physics as a discipline ``$\dots$ full of mysteries and paradoxes, that we do not completely understand, but are able to use. As we know, it perfectly operates 
in the physics reality description, but as sociologists would say, it is an
anti-intuitive discipline. The quantum physics is not a theory, but limits, 
in which, as we suppose, any correct theory needs to be included''. Now, we can clearly see it from our point of view. \\

For the further analysis it is important to mention that the conclusions in this section are valid not only for the axis space rotation, but also for any rotation, where the $\omega$-invariance principle may be introduced. For example, it needs to be valid for any MSR, if all ASRs, included in this MSR, have equal or multiple frequencies and are normalized. The requirement of the normalization is coming from (\ref{n7}). In case of the SSR, this requirement is unable to satisfy due to SSR transformation matrix properties (Sect. \ref{sec:Types}). This condition can not be satisfied principally, because the functional determinant of the SSR transformation matrix is time-dependent. 

\section{Space Rotations and Metrics}
\label{sec:Metrics}

It was mentioned before in Section \ref{sec:Types}, that the space rotation invariance, if it exists (here we do not mean the $\omega$-invariance), has a local character, so we will analyze the SR metrics from some fixed space point in the initial $K$ and rotating $K'$ frames. We will call this point as a ``rotation point''. The space coordinates of this point in $K$ and $K'$ are $(0,0,0)$. So, from this point we will use the following assignments for the spherical ($r^2=\, x^2+ y^2+ z^2,~r\, \cos(\theta)=\, z,~x\, \tan(\phi)=\, y$) and cylindrical ($\rho^2=\, x^2+ y^2,~z=\, z,~\rho\, \cos(\phi)=\, x$) coordinates. We will use different coordinates whenever it will be easier to make an analysis. \\

Note, that the metrics from the rotation point for the ASR and MSR (as it is understood in the Minkowski space) in $K$ and $K'$ is the same. Indeed, if we will consider the interval from the rotation point $X'_{rp}=\, X_{rp}=\, (0,0,0)$ (the same in $K$ and $K'$) at zero time $t=\, 0$ to some space point at time $t$, according to (\ref{n5}), we will get for any time $t$ ($X'=\, X'_2,\, X=\, X_2,\, X'_1=\, X_1=\, X_{rp},\, t_2=\, t,\, t_1=\, 0$): 

\begin{equation}
\|X'\|^{2}=\, X'\cdot X'^{T}=\, X\cdot A^{MSR}(t)[A^{MSR}(t)]^{T}\cdot X^{T}=\, 
X\cdot X^{T}=\, \|X\|^{2}, 
\label{XN}
\end{equation}
 
due to the ASR and MSR transformation matrix properties (see Sect. \ref{sec:Types}). This really means that the space rotation invariance has a local (or even point) character. The situation with points differed from the rotation point is another.\\ 

Let's consider some space rotation with the transformation matrix $A$ between $K$ and $K'$.
Between time points of $\omega$-invariance for space points differed from the rotation point the equality (\ref{XN}) is not true. We will try to find the interval in the differential form. The differential $dX'$ will be defined as: 

\begin{equation}
dX'=\, d(X\cdot A)=\, d\left \{ \sum_{i} x_{i}a_{ij}\right \} =\, \left \{ \sum_{i} dx_{i} a_{ij}\right \}+\left \{ \sum_{i} x_{i} da_{ij}\right \} =\, dX\cdot A + X\cdot dA. 
\label{n8}
\end{equation}

For the interval $ds'^{2}=\, c^{2}dt^{2}-\, \|dX'\|^{2}$, where  $\| dX'\|^{2}=\, dX'\cdot dX'^{T}$, using (\ref{n8}), we will get: 

\begin{equation}
\left (ds'_{SR}\right )^{2}=\, c^{2}dt^{2}-\, \| dX'\|^{2}=\, c^{2}dt^{2}-\, X\, dA\, dA^{T}\, X^{T}-\,  dX\, A\, A^{T}\, dX^{T}-\, \left \{ dX\, A\, dA^{T}\, X^{T}+\, X\, dA\, A^{T}\, dX^{T} \right \}. 
\label{X2}
\end{equation}

This expression is true for any space rotation. We will analyze the ASR and MSR with help of the expression: 
 
\begin{equation}
\left (ds'_{MSR}\right )^{2}=\, \left [ c^{2}-\, \left ( X\, \frac {\partial A}{\partial t}\, \frac {\partial A^{T}}{\partial t}\, X^{T}\right )\right ]\, dt^{2}-\,  \|dX\|^{2}-\, \left \{ dX\, A\, dA^{T}\, X^{T}+\, X\, dA\, A^{T}\, dX^{T} \right \},  
\label{X}
\end{equation}

that follows from the expression (\ref{X2}) by using the MSR properties (see Sect. \ref{sec:Types}).  
This expression defines the ''real'' metrics for the ASR or MSR in $K'$ from $K$ point of view and the difference between intervals in $K$ and $K'$ may be observed clearly. \\

{\bf The ASR metrics.} Let's consider the transformation matrix $A^{+}_{z}$ from (\ref{ASR}) without loosing the generality. From (\ref{X}), we will get the equation: 

\begin{equation}
\left (ds'_{ASR}\right )^{2}=\, \left [\, c^2-\, \omega^{2} \left (x^{2}+y^{2}\right )\right ]\, dt^{2}-2\,\left (y\, dx-\,x\,dy\right )\omega\, dt- (\, dx^{2}+dy^{2}+dz^{2} ).
\label{dX2}
\end{equation} 

It is easy to analyze this expression in cylindrical coordinates in $K$: 

\begin{equation}
\left (ds'_{ASR}\right )^{2}=\, \left (c^2-{\omega}^{2}\rho^{2}\right ){{\it dt}}^{2}+\, 2\rho^{2}\omega\, d\phi\, dt-\, \rho^{2}\, d\phi^{2}-{{\it d\rho}}^{2}-{{\it dz}}^{2}. 
\label{ASRint}
\end{equation}

It is seen from here, that the interval $ds'^{2}$ is not invariant from usual point of view, the term in brackets at $dt^{2}$ depends on space coordinates and even changes the sign in some space points. If we are moving radially, it is seen, that this term is positive when $\rho^{2}<\, c^2/{\omega}^{2}$ and is negative when $\rho^{2}>\, c^2/{\omega}^{2}$. It means that metrics is changed and {\it the functional determinant of the metrics tensor changes the sign} in these regions, so the space-time in $K$ becomes like a four-dimensional Euclidean space in the ``external'' regions. \\ 

It is enough for our analysis, that there exist some regions in space in $K'$ from the point of veiw $K$, where the coefficient at ${{\it dt}}^{2}$ (\ref{ASRint}) is equal to zero:

\begin{equation}
\rho^{2}=\, c^2/{\omega}^{2}.
\label{ASRobj}
\end{equation}

It means that there are some regions localized in $x0y$-plane in space, where the physical object, whenever it is placed there, will be stable in time from the $K$ point of view! It is seen from (\ref{ASRobj}) that for the ASR the stable region is cylindrical around the rotation axis. The metrics along the $z$-axis is like in Minkowski space. The ASR object is shown on Fig. \ref{fig1}. \\ 

If there exists the physical object in $K'$, from $K$ frame we will observe the object, localized in $x0y$-plane and moving along with the $z$-axis. As far as the rotation frequency with the same value may have two direction of rotation (may have positive or negative sign), it may be two different types of this objects, or one can say that this object will have an additional characteristic from the $K$ point of view. These characteristic looks like {\it spin}, that on this approach may be directed along or opposite the $z$-axis. This model of the ASR physical object has some correspondence with {\it neutrino}. \\

{\bf The MSR metrics.} We will consider the MSR, consisted of three orthogonal ASRs with rotation frequencies $\omega_{1},\, \omega_{2}$ and $\omega_{3}$, so $\, A^{MSR}=\, A^{ASR}_{z}(\omega_{1})\, A^{ASR}_{x}(\omega_{2})\, A^{ASR}_{y}(\omega_{3})$. For the MSR the analysis is differed from the ASR one, because the interval (\ref{X}) depends on time and is complicated. For example, for $A^{ex}=\, A^{ASR}_{z}(\omega)\, A^{ASR}_{x}(\omega)$, even $\|dX'\|^{2}$ will be expressed as: 

\begin{eqnarray}
\|dX'\|^{2}  =  \left [2\,{x}^{2}-2\,x\sin(\omega\,t)y\cos(\omega\,t)+{y}^{2}-{x}^{2}
\cos^{2}(\omega\,t)-2\,yz\sin(\omega\,t)+{y}^{2}
\cos^{2}(\omega\,t)-2\,xz\cos(\omega\,t)+{z}^{2}\right ]{
\omega}^{2}{{\it dt}}^{2}+\nonumber\\ 
+2\, \left [\left (y-\,z\sin(\omega\,t)
\right ){\it dx}+\left (z\cos(\omega\,t)-\,x\right ){\it dy}-
\left (\cos(\omega\,t)y-\,\sin(\omega\,t)x\right ){\it dz}\right 
]\omega\,{\it dt}+{{\it dx}}^{2}+{{\it dy}}^{2}+{{\it dz}}^{2}. \nonumber  
\end{eqnarray}

We will analyze the interval, averaged in time, from the point of view of the observer in $K$. We will consider that the period of the time of the observation is much longer than the period of any rotation included in the MSR. We will use the expression for ``averaging'': 

\begin{equation}
\langle f \rangle_{t}=\, \lim _{t\to \infty}\frac {1}{2t}\int_{-t}^{+t}\, f(t)\, dt.
\label{average}
\end{equation} 

This approach simplifies expressions, so we can analyze them. For example, the interval for $A^{ex}$ will be: 

\begin{equation}
\langle \left (ds'_{ex}\right )^{2} \rangle_{t}=\, \left [c^{2}- \left (\frac {3}{2}\,{x}^{2}+\, \frac {3}{2}\,{y}^{2}+{z}^{2}\right ){\omega}^{2}\right ]{{\it dt}}
^{2}-2\, \left (y\,{\it dx}-\,x\,{\it dy}\right )\omega\,{\it dt}-{{
\it dx}}^{2}-{{\it dy}}^{2}-{{\it dz}}^{2}.\nonumber
\end{equation} 

On this way, for the MSR, consisted of three orthogonal ASRs with rotation frequencies $\omega_{1},\, \omega_{2}$ and $\omega_{3}$, so $A^{MSR}=\, A^{ASR}_{z}(\omega_{1})\, A^{ASR}_{x}(\omega_{2})\, A^{ASR}_{y}(\omega_{3})$, one can get: 

\begin{eqnarray}
\langle \left (ds'_{MSR}\right )^{2} \rangle_{t}=\, \left \{ c^2-\, \left [\left ({x}^{2}+{y}^{2}\right ){\omega_{1}}^{2}+\left (\frac {1}{2}\,{y}^{2}+
{z}^{2}+\, \frac {1}{2}\,{x}^{2}\right ){\omega_{2}}^{2}+\left (\frac {1}{2}\,{z}^{2}+\frac {3}{4}\,{y}^{2}+\frac {3}{4}\,{x}^{2}\right ){\omega_{3}}^{2}\right ]\right \}{{\it dt}}^{2}- \nonumber \\ 
-2\,\left (y\,{\it dx}-\,x\,{\it dy}\right ){\omega_{1}}\,{\it dt}-{{\it dx
}}^{2}-{{\it dz}}^{2}-{{\it dy}}^{2}.
\label{MSRint}
\end{eqnarray} 

In cylindrical coordinates, this equation can be rewritten as: 

\begin{equation}
\langle \left (ds'_{MSR}\right )^{2} \rangle_{t}=\, \left \{ c^2-\, \left [\left (\rho^{2}\right ){\omega_{1}}^{2}+\left (\frac {1}{2}\,\rho^{2}+
{z}^{2}\right ){\omega_{2}}^{2}+\left (\frac {1}{2}\,{z}^{2}+\frac {3}{4}
\,\rho^{2}\right ){\omega_{3}}^{2}\right ]\right \}{{\it dt}}^{2}+\, 2\rho^{2}\omega\, d\phi\, dt-\, \rho^{2}\, d\phi^{2}- d\rho^{2}-{{\it dz}}^{2}.
\label{Mrint}
\end{equation} 

On the analogy of the ASR analysis, one can conclude that there are some regions localized in space, where the physical object, whenever it is placed there, will be stable in time from the $K$ point of view! In spherical coordinates these stable regions will satisfy the equation: 

\begin{equation}
r^{2}=\, \left [\left (\frac {\omega_{1}}{c}\right )^{2}\, \sin^{2}(\theta)+\, 
\left (\frac {\omega_{2}}{c}\right )^{2}\,\left (1-\,\frac {1}{2}  \sin^{2}(\theta)\right )+\,
\frac {1}{2}\left (\frac {\omega_{3}}{c}\right )^{2}\,\left (1+\,\frac {1}{2}  \sin^{2}(\theta)\right )\right ]^{-1}.
\label{MSRobj}
\end{equation} 

This equation describes an ellipsoid in space. The ellipsoidal MSR object for $A^{ex}$ is shown on Fig. \ref{fig2}. The equation (\ref{ASRobj}) is described by the equation (\ref{MSRobj}) with the assignment $\omega_{1}=\, \omega,\, \omega_{2}=\, \omega_{3}=\, 0$. This way it will represent the cylindroid around the $z$-axis. \\  

Comparing the expression (\ref{MSRint}) with (\ref{ASRint}), one can see that although the MSR consists of few orthogonal ASRs, a very important role plays the ASR, corresponding to the first transformation matrix of the MSR transformation matrices set. In our case, it is the ASR, corresponding to rotation about $z$-axis with the transformation matrix $A^{ASR}_{z}(\omega_{1})$. The MSR object, localized in space, anyway has some axis, picked out in space. We can suppose, as before, during the ASR object analysis, that it means that the MSR object also needs to have the physical ``spin''-characteristic. Such SR objects are similar to {\it fermions}. \\ 

{\bf The SSR metrics.} We will consider the SSR, consisted of three orthogonal ASRs with rotation frequencies $\omega_{1},\, \omega_{2}$ and $\omega_{3}$, so $A^{SSR}=\, A^{ASR}_{z}(\omega_{1})+\, A^{ASR}_{x}(\omega_{2})+\, A^{ASR}_{y}(\omega_{3})$. For the SSR, we need to analyze the SSR metrics by using the expression (\ref{X2}). Analogous to the MSR metrics analysis, instead of (\ref{MSRint}), for the SSR interval, one can get: 

\begin{eqnarray}
\langle \left ( ds'_{SSR}\right )^{2} \rangle_{t}=\, \left \{ c^2-\, \left [\left ({x}^{2}+{y}^{2}\right )\omega_{1}^{2}+\left ({y}^{2}+{z}^{
2}\right )\omega_{2}^{2}+\left ({x}^{2}+{z}^{2}\right )\omega_{3}^{2}\right ]\right \}{{\it dt}}^{2}-\nonumber\\ 
-2\, \left [\left (y\,{\it dx}-x\,{\it dy}\right )\omega_{1}-\left (y\,{\it dz}+\,z\,{\it dy}\right )
\omega_{2}+\left (z\,{\it dx}-\,x\,{\it dz}\right )\omega_{3}
\right ]{\it dt}-\nonumber\\ 
-3\,\left ({{\it dx}}^{2}+\,{{\it dy}}^{2}+\,{{\it dz}}^{2}\right ).
\label{SSRint}
\end{eqnarray} 

Here, we can see that the SSR interval in average is equal to a sum of the intervals of included orthogonal ASRs. By analogy with the analysis of the MSR metrics, we can conclude that the SSR object is localized in space. The SSR object has an ellipsoidal structure. But it is impossible to pick out any space axis for this object as a whole. The ``space distance'' in $K$ ($\|X\|^{2}$) is multiplied by some coefficient (it is $3$ in our case). It is not a matter of calibration, because the considered ASR transformation matrices, included in the SSR, are normalized and so the scale from the ASRs point of view is differ from these ASRs point of view ``as a whole''. All conclusions in this paragraph are valid in case of the SSR consisted of two orthogonal ASRs. As far as we have considered that the MSR object has some picked out space axis, all conclusions are valid in case of the SSR consisted of the orthogonal MSRs, of course, in frames of limitations due to ``averaging''. \\ 

On this approach, if we will continue the analogies between SR objects and particles, it is possible to find some correspondence of the SSR objects with models of the particles consisted of {\it quarks} (the SSR of ASR objects) and even with the {\it nuclei} (the SSR of MSR objects). \\

It needs to be mentioned, that it is impossible to normalize the SSR transformation matrix  even in ``averaging''. For the SSR objects, the equality (\ref{n7}) is not true (with ``averaging'' some additional coefficient will appear), and so the quantum mechanics equations of Sect. \ref{Quantum} for SSR objects need to be modified. It looks like due to the SSR metrics changing the intensity of the ``interaction'' between the included ASRs is increasing, that, may be, will correspond to electroweak and strong forces. It has some analogies with the chromodynamics. On our approach, we do not need to postulate some additional objects and forces - they are included into the space rotation invariance hypothesis. We can know for sure the limits of the implementation of our theory, its correspondence to the classical electrodynamics and relativistic theories. \\ 

We need to mention once more, that we analyzed the MSR and SSR objects by averaging the interval in time, the complete picture is more complicated, but it is not a principal limitation, only some type of approximation. We will leave the complete analysis of these objects for the future. \\

\section{Principle of Quantification}
\label{sec:Quant}

In this section we will investigate the question of the possibility of the ``existence'' of the SR objects in the observer frame of references $K$ from the point of view of known physics laws in this frame. We will consider the case when the SR object in $K'$ is a source of some ``influence'' in $K$: a field, a force or something else. We consider, that this ``influence'' is well known in $K$ (here, we don't mean quantum physics). One knows, for example, that the charged particle, moving in space is a source of the electromagnetic radiation. If the influence takes some energy from the source, we need to require for a stable object that somewhere in space around the object, this influence or, in other words, the energy flow from the object is zero. So, if the influence in $K$ from the SR object exists, it has to exist only in some limited region around the SR object, i.e. this influence needs to have a local character. \\ 

First, we will consider the model or an ``idea'' of such influence in one-dimensional space. We will consider the following model: there are two sources of ``influence'' on the $x$-axis in point $-a$ and $a$. The influence with some characteristic function $u(x,t)$ spread with the speed $v$ from the source in both directions - along the $x$-axis ($u(x-vt)$) and in the opposite direction ($u(x+vt)$). We will consider that the superposition for this influence is valid. \\ 

So, these two sources separate the $x$-axis on three intervals: $(-\infty,\, -a),\, (-a,\, a)$ and $(a,\, +\infty)$. For the influence characteristic function $u(x,t)$ we can write: 

\begin{equation}
u(x,t)=\left\{ \begin{array}{ll} 
u\left [(x+a)-vt\right ]+\, u\left [(x-a)-vt\right ] & \textrm{ for $(a,\, +\infty),$} \nonumber\\
u\left [(x+a)-vt\right ]+\, u\left [(x-a)+vt\right ] & \textrm{ for $(-a,\, a),$} \nonumber\\ 
u\left [(x+a)+vt\right ]+\, u\left [(x-a)+vt\right ] & \textrm{ for $(-\infty,\, -a).$} \nonumber \\
\end{array} \right. 
\end{equation}

These requirements can be satisfied, if, for example, there is a source in one point $u(x,t)>\, 0$ and a drain in another $u(x,t)<\, 0$. Also, if the function $u(x,t)$ is periodical with period $2\pi$: $u(\xi)=\, -u(\xi+\, \pi m)=\, u(\xi+\, 2\pi m)$, $m$ is integer, one can get the zero influence in ``external'' regions $(-\infty,\, -a)$ and $(a,\, +\infty)$. There are two decisions: the ``symmetrical'', when $a=\, \pi/2+\pi m$ and ``anti-symmetrical'' one with $a=\, \pi m$. It is possible to consider the drain-source model as a particular case of the periodical function model with the infinite period. \\

In other particular case, the influence is described by the periodical wave function $u(\xi)=\, c_{1}\, \cos(\xi)+\, c_{2}\, \sin(\xi)$. The symmetrical solution (even mode) corresponds to $c_{2}=\, 0$, and anti-symmetrical one (odd mode) corresponds to $c_{1}=\, 0$. \\

In three-dimensional space, it is possible to set the problem as follows. If the stable SR object exists in some limited space region $G$ and this object is the source of the wave-like field $u(X,t)$, so it needs to satisfy the wave equation: 

\begin{equation}
\nabla^{2}\, u(X,t)+\, \frac {1}{v^{2}}\,\frac {\partial^{2}}{\partial t^{2}} u(X,t)=\, -f(X,t), 
\label{waveeq}
\end{equation} 

where $v$ is a speed of the influence wave, $f(X,t)$ is some source function. We consider the SR object as a source of the influence field $u(X,t)$, so, following Sect. \ref{sec:Quantum} (\ref{ps}) in frames of $\omega$-invariance we can put for source function: 

\begin{equation}
f(X,t)=\, q(X)\, \exp(\pm {\bf i}\, \omega t).
\label{source}
\end{equation}

Usually, in transmission from quantum mechanics to electrodynamics, the source function is defined as $q(X)=\, \|\psi(X,t)\|^{2}$. It is not satisfactory for us, because this way the internal, source characteristics are neglected, but there are these properties we want to investigate. \\ 

After that, seaching the solution in form $u(X,t)=\, U(X)\, \exp(\pm {\bf i}\, \omega t)$, we wil come directly to a well known differential equation for function $U(X)$ \cite{Vl}: 

\begin{equation}
\nabla^{2}U(X)+\, k^{2}U(X)=\, -q(X).
\label{DE}
\end{equation}

Here $k=\, \omega/c$. 
To find the solution, one needs to know the concrete function q(X). It is not necessary for our analysis. We need to require for a stable object that somewhere in space around the object, this influence or, in other words, the energy flow from the object is zero. We will replace the equation (\ref{DE}) and consider also well known equation for the external sphere potential \cite{Vl}. In spherical coordinates: 

\begin{eqnarray}
\nabla^{2}U(r,\theta, \phi) & + & \, k^{2}U(r,\theta, \phi)=\, 0, \label{Sph}\\
U_{0}(\theta, \phi) & = & \, \sum_{l=0}^{\infty}\sum_{m=-l}^{l}\, a_{lm}Y^{m}_{l}(\theta,\phi), \nonumber
\end{eqnarray} 

where $Y^{m}_{l}(\theta,\phi)$ are spherical functions and $U_{0}(\theta, \phi)$ is defined on some sphere $S_{R}$ of radius $R$. The solution of this equation will be: 

\begin{eqnarray}
U(r,\theta,\phi) & = & \, \sum_{l=0}^{\infty}\sum_{m=-l}^{l}\, 
R_{lm}(r)Y^{m}_{l}(\theta,\phi), \label{Sol}\\
R(r) & = & \, \frac {C_{1}}{\sqrt {r}}J^{(1)}_{l+\frac {1}{2}}(kr)+\, \frac {C_{2}}{\sqrt {r}}J^{(2)}_{l+\frac {1}{2}}(kr), \nonumber
\end{eqnarray} 

where $C_{i}$ are some constants, $J^{(i)}_{l+\frac {1}{2}}(kr)$ are Bessel's functions of the first and the second kind correspondingly. As far as, the function $U(X)$ needs to be zero at $S_{R}$, so $U(R)=\, U_{0}=\, 0$, from (\ref{Sph}) and (\ref{Sol}), we will get the necessary condition for the stable SR object existence in $K$: 

\begin{equation}
\left [
\begin{array}{ll} 
J^{(1)}_{l+\frac {1}{2}}(kr)=\, 0, & \textrm{for even mode,} \\
J^{(2)}_{l+\frac {1}{2}}(kr)=\, 0, & \textrm{for odd mode.} \\ 
\end{array} \right.
\label{QP}
\end{equation}

So, the SR objects in $K$ are allowed to have discrete `` sizes'' $r_{i}=\, \alpha_{i}/k$, where $\alpha_{i}$ are the zeros of the Bessel functions. We will call these necessary conditions as a {\it Quantification principle}. The quantification takes place without any ``external'' force field. We need to mention, that there are no limitation of the speed of the SR objects in $K'$ inside ``limited'' regions of their existence from the observer in $K$ point of view. But we do not have any reason to consider that the speed of these objects in $K'$ is more than the speed of light.    

\section*{conclusion}

It was found during the investigation an impressive correspondence of the consequences of the space rotation invariance hypothesis to basic postulates of the quantum physics.\\ 

The principle of the space rotation invariance, declared in this paper, means physics law invariance together with the physical objects in different space rotation frames of references. It does not mean the interval invariance as it was declared in special theory of relativity. The space rotation invariance is different also from the Einstein's general relativity theory. The space rotation invariance has a local character, but seems to be more fundamental than these invariance principles. \\ 

The particular case of SR invariance is an $\omega$-invariance also introduced in this paper. The $\omega$-invariance is defined on the numerable aggregate on the time-axis. On this case, it is shown, that it is possible to get the basic quantum mechanics equations, such as Schr{\"o}dinger and Klein-Gordon-Fock, for the description of the SR objects. On this approach, it becomes clear that the $\omega$ - invariance is the reason of the {\it uncertainties} in quantum mechanics, the reason of its incompleteness and formalism. Anyway, it makes clear the limits of the quantum mechanics formalism and quantum mechanics objects. A lot of quantum mechanics ``paradoxes'', such as a particle dualism, a wave function collapse, are getting quite clear physical explanations. \\ 

The correspondence of the properties of the different SR objects with some properties of the elementary particles, quarks and even nuclei, makes to hope that the space invariance hypothesis, introduced in this paper, will grow to some unification theory. \\ 

May be, it will be possible to include also a gravitation in this future unification theory by analyzing the ``external'' regions of SR objects with the four-dimensional Euclidean-like structure. 

% now the references. delete or change fake bibitem. delete next three
%   lines and directly read in your .bbl file if you use bibtex.

% \begin{figure}
% \caption{}
% \label{}
% \end{figure}
%
\begin{figure}
\caption{The ASR object.}
\label{fig1}
\end{figure}

\begin{figure}
\caption{The MSR object.}
\label{fig2}
\end{figure}

% \begin{table}
% \caption{}
% \label{}
% \end{tabular}
% \end{table}

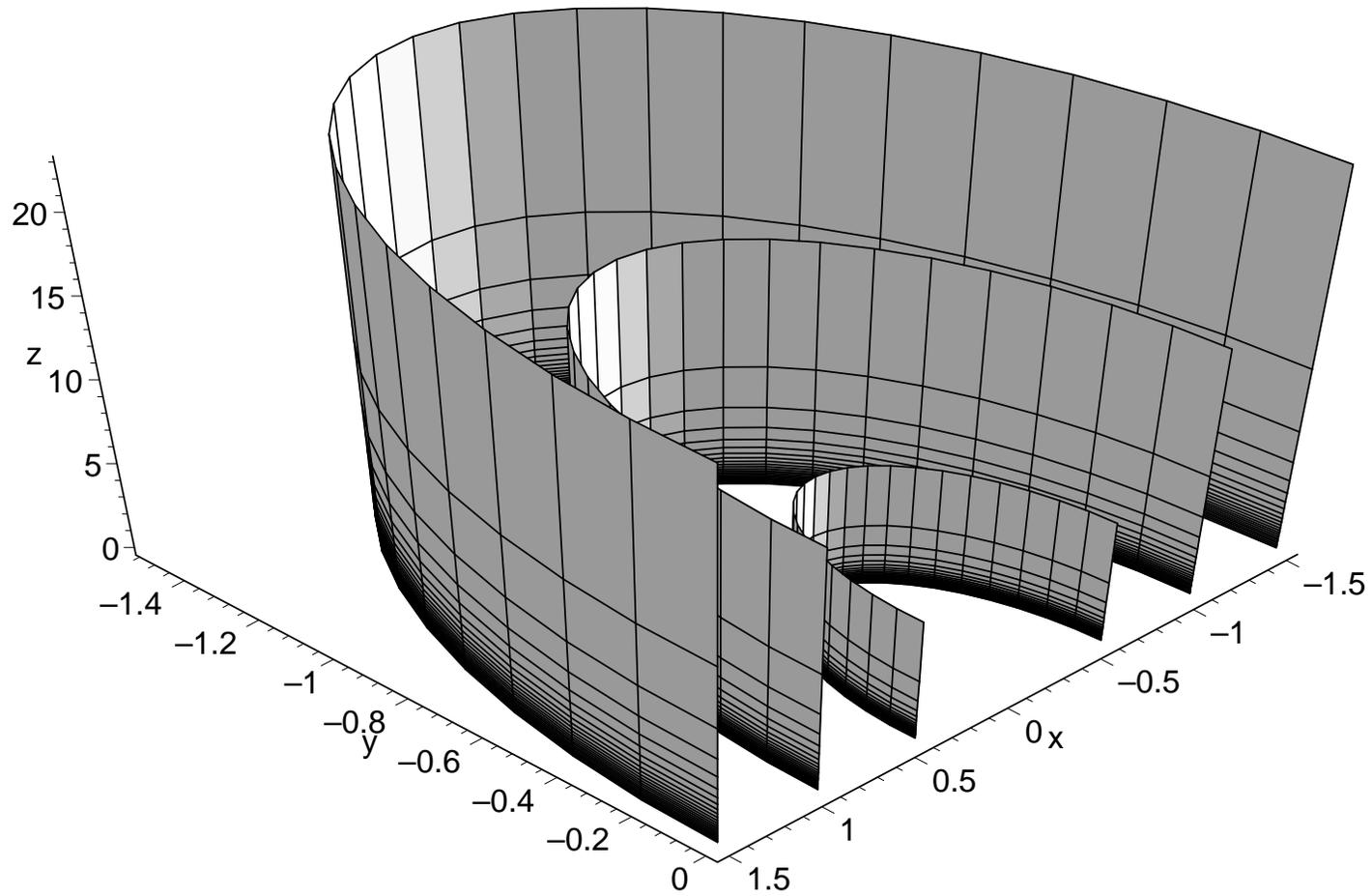
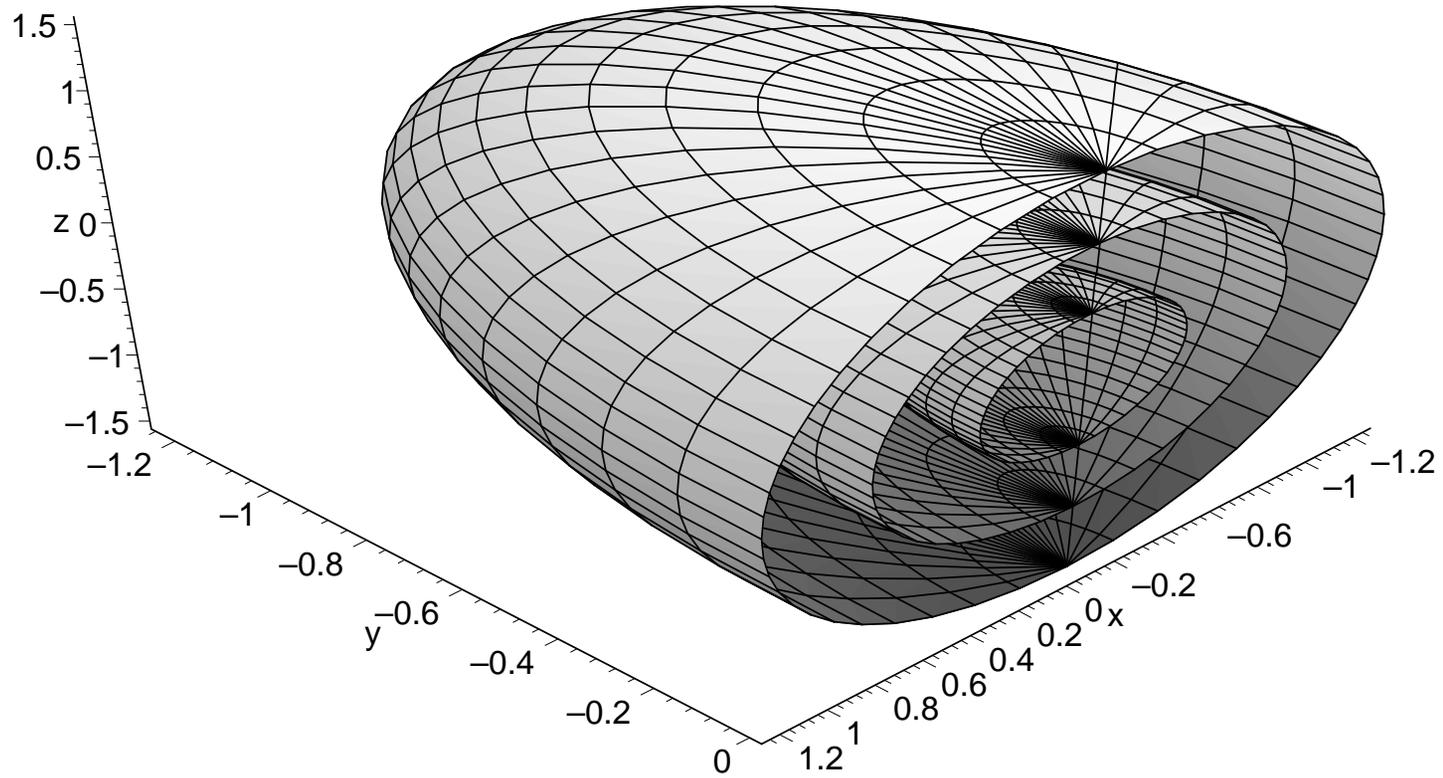
\end{document}